\newenvironment{namelist}[1]{%
\begin{list}{}
    {
      
      \settowidth{\labelwidth}{#1}
      \setlength{\leftmargin}{1.1\labelwidth}
    }
  }{%
\end{list}}

\documentstyle[epsf,preprint,twoside,adbis97%
,amscd%
,times]{acmconf} 

\author{V.E.Wolfengagen \vspace{1.52mm} \\
{Vorotnikovsky per., 7, bld. 4} \\
{Institute for Contemporary Education ``JurInfoR-MSU''} \\
{Moscow, 103006, Russia} \\
{\tt vew@jmsuice.msk.ru}
}
\addtocounter{footnote}{1}
\title{Building the access pointers to a computation environment
\thanks{\em This research is supported by the
Russian Foundation for Basic Research
(project 96-01-01923)}\\%
}
\newtheorem{lem}{\sf Lemma}[section]
\newtheorem{th}{\sf Theorem}[section]

\begin{document}
%%%%%%%%%%%%%%%%%%%%%% Bibliographystyle  %%%%%%%%%%%%%%%%%%%%%%%%
\bibliographystyle{alpha}
%%%%%%%%%%%%%%%%%%%%%%%%%%%%%%%%%%%%

\maketitle

%%%%%%%%%%%%%%%%%%%%% Abstract %%%%%%%%%%%%%%%%%%%%%%%%%

\begin{abstract}
%This extended abstract covers some selected questions
%from the entire paper.
A common object technique equipped with the categorical and computational
styles is briefly outlined.
An object is evaluated by
embedding in a host computational
environment which is the domain-ranged structure.
An embedded object is accessed by the pointers generated within
the host system.
To assist with an easy extract the result of the evaluation
a pre-embedded object is generated.
It is observed as
the decomposition into substitutional part and access function
part which are generated during the object evaluation.
\end{abstract}

%%%%%%%%%%%%%%%%%%%%%%%%%%%%%%%%%%%%%%%%%%%%%%%%%%%%%%%

%%%%%%%%%%%%%%%%%%%%%%%%%%%%%%%%%%%%%%%%%%%%%%%%%%%%%
%%%%%%%%%%% The body of the contribution %%%%%%%%%%%%
%%%%%%%%%%%%%%%%%%%%%%%%%%%%%%%%%%%%%%%%%%%%%%%%%%%%%

\section{Introduction}
% \addcontentsline{toc}{section}{Introduction}
Recent issues in a data modeling area tend to
attract some general algebraic ideas.
The most influent to the target data model
are the properties of the database domains
and their interconnections.

Some useful observations concerning the mappings
between the {\em relational database} domains
\cite{Article:94:Islam:CatTheoRM} result in
the solutions to integrate a database scheme.
But the difficulties were observed when the
{\em type} considerations occurs: the first-order
data model becomes overloaded with the complicated
and intuitively unreasonable mappings, especially
when attempts to use a {\em category} theory are done.

The attempts to apply the same ideas for
a {\em conceptual modeling} \cite{Report:94:Lippe:CatTheory}
are not yet advanced to cover the known effects
and models. A gap between the pure reasoning with
the objects in a category-style manner
(the maps and domains have the similar status)
and the realistic data models is indicated
every time (see, e.g., \cite{Report:94:Hofstede:CTPerspective})
when the researcher put the database concepts together.
Nevertheless, the feeling of a category theory usefulness
is growing with the rate of accumulating  the practical
experience in a field \cite{PhdThesis:91:Jacobs:CategoryTheory}.

The {\em semantics} of database is heavy based on the evaluation
of the expressions \cite{Article:94:Baclawski:CatTheoDatabase}.
The success of the approach is also estimated by the simplicity
and intuitive transparency whenever the maps between domains are
involved. Information system engineering
\cite{Article:93:Johnson:AMAST93} extremely needs to apply the
theoretically balanced data models with the higher-order
structures.

The observations show the concepts and notions shared by the
distinct approaches and the theories. The importance of extracting
all the useful feature from the notions of {\em function} and {\em
type} are well understood. Here we will try to rearrange and put
together some important ideas concerning the evaluation of
expressions. The most of the attention is paid to {\em
environment} of the evaluation to suit it with the common database
models. An environment is assumed to consist of the {\em products}
of the domains. Thus, the obvious way to get an access to its
partitions is to evaluate the {\em projections}, each of them
being type according to the positions of the counterparts. This
intuitively means, the {\em pointers} to an environment are to be
generated. Some unexpected features arise wherever the {\em
encapsulation} of objects is used relatively the environment
prescribed.

To cover the notion in use the most of attention is paid
to interrelations and correspondences  between
{\em types}, {\em functions} and {\em environment}
of evaluation. The {\em language} is left out
of this paper scope, and model theoretic aspects
are attracted. The style of reasoning in a category
is used along with the equational solutions.

Section~\ref{Section:1} covers the minimal amount of a type theory
to put the necessary accents. The projections are used almost in a
traditional sense. The operation $\hat{}$ gives a kind of suit to
shift around the variables. In addition, the correspondence
between the projections and the {\em identity maps} brings in a
theory the intuitive ground.

%In section~\ref{Section:2} the general considerations
%are given.

An access to values coupled in environment is discussed
in section~\ref{Section:3}. The main topics are generation
of the {\em access pointers} and the {\em encapsulation}
of the objects. The commutative diagram techniques
is applied to establish the most important equation.
The reasons and solutions are based on the possibilities
of the {\em citation}. An atomic case is covered by Lemma~\ref{Lemma:1}.
Its generalization leads to the Theorem~\ref{Theorem:1}
whenever the function constant is applied to an argument.
To pass an {\em actual parameter} the {\em closure}
is generated.

\section{A theory of types}\label{Section:1}

A variety of possible theories of types has been developed
with different purposes and with distinct mathematical or logical
ideas in use. A category theory gives one of the theories.

To establish a universe of discourse for types we need
some kind of {\em primitive frame}. Usually we start with
a set of {\em generic} types and generate the  {\em derived}
types applying some building principles. In a pure category
theory we are not given neither building principles nor clear
understanding of making new types from old. Entering a category
theory we observe the relations between types which hold
whenever corresponding mapping statement
\begin{equation}
%$$
f: Env \to D_y %\eqno{(f)}
%$$
\end{equation}
obtains (here: $Env$ and $D_y$ are domains).

The way of reading the {\em mapping statement} depends on
intuitive reasons. Whenever we apply to category theory,
a mapping statement is supposed being taken as a statement
with one-place functions and operation `$\circ$' of composition
with one-place functions. Thus, practical reasons concern
the multi-place functions to increase their arity.

The solutions to bring composition with multi-place functions
(see, e.g., \cite{Sza:78}) are known but give no real suit.

The easier way is to assume that the category has
{\em cartesian products} and to select the particular
representatives of the product domains. In particular,
the cartesian power $D^n$ for every $n \ge 0$ gives $n$-ary
functions as maps
\begin{equation}
f: D^n \to D_y.
\end{equation}

\subsection{A description of products}

\subsubsection{Empty product.}

To make a description of products we bring in the {\empty} product
and start with the assumption that a category has a special domain
$\cal O$ as the empty product:
\begin{equation}
D^0_y = D^0_x = \cal O,
\end{equation}
and for every domain $D$ a special map:
\begin{equation}
0_D: D \to \cal O
\end{equation}
From an intuitive reason the domain ${\cal O}$ has one
element, and map $0_D$ is unique, i.e. whenever
$f: D \to {\cal O}$ then $f = 0_D$. \\[1ex]

\subsubsection{A theory of tuples and multi-ary maps.}

This kind of a theory
is based on the {\em products}. Concerning {\em binary products}
we have for arbitrary two domains $D_x$ and $D_y$ a special choice
of a domain $D_x \times D_y$, and, more generally,
$$
D^{n+1} = D^n \times D,\ n > 0.
$$
A product is equipped with the special maps
\begin{eqnarray*}
Fst & : & D_y \times D_x \to D_y, \\
Snd & : & D_y \times D_x \to D_x, \\
\end{eqnarray*}
which are the {\em projections}. As usually, mere existence of
maps $Fst$ and $Snd$ does not characterize $D_y \times D_x$
as a product. In addition we assume that there is a chosen
{\em pairing operation} $<f,g>$ on maps such that types are
assigned by the rule:
\[
\begin{array}{c}
f: Env \to D_y,\ \ g: Env \to D_x \\
\rule[0pt]{14em}{0.05em} \\
<f,g>: Env \to D_y \times D_x
\end{array}
\]
The additional property of $Fst,\ Snd$ and $<\cdot,\ \cdot>$
under {\em composition} is assumed:
\[
\begin{array}{rcl}
Fst \circ <f, g> & = & f, \\
Snd \circ <f, g> & = & g, \\
<Fst \circ h, Snd \circ h> & = & h,
\end{array}
\]
where $f,\ g$ are typed as above, and
$$
h: Env \to D_y \times D_x
$$
\begin{namelist}{{\tt m} {\it {  }}}
\item {\hspace{-1em}\sf One-to-one correspondence.}
It means that there is a one-one correspondence
between the pairs of maps $f,\ g$ and the map $h$
into the product.
\end{namelist}

\subsubsection{A theory of functions.}

A category usually gives a `local' universe of {\em selected}
functions. In case of {\em arbitrary} functions we need the {\em
functional spaces} as explicit domains in the category.

Given $D_x$ and $D_y$ we want to form $(D_x \to D_y)$
as a domain in its own right. After adopting the above,
the functional space does contain the various maps.

Whenever we have an element $f$ from $(D_x \to D_y)$ and the
element $x$ from $D_x$ we need to establish the map that
will apply element $f$ to element $x$ giving rise to the
{\em value} of function $f$:
$$
\varepsilon : [f,x] \mapsto f(x)
$$
This {\em evaluation} map $\varepsilon$ is typed as
$$
\varepsilon : (D_x \to D'_y) \times D_x \to D'_y
$$
In addition there has to be a map for shifting around
{\em variables}. Suppose
$$
g: Env \times D_x \to D'_y
$$
is a map with two arguments. In an evaluation
$$
g([i,x])
$$
we can think of holding $i$ constant and regarding $g([i, x])$
as a function of $x$. We need a name for this function
and for correspondence with possible values of $x$:
$$
{\hat g}: Env \to (D_x \to D'_y)
$$
so that the function we are thinking of - given $x$ - was
$$
{\hat g}(i)(x)
$$
\begin{namelist}{{\tt -h} {\it {   }}}
\item {\hspace{-1em}\sf Map $k$ is one-to-one corresponded to $g$.}
All this function value notation
is not categorical notation. Nevertheless we are to say
that there is a one-one correspondence via $\hat \cdot$
between maps $g: Env \times D_x \to D_y$
and maps $k: Env \to (D_x \to D_y)$.
\end{namelist}
This correspondence comes down to the following two
equations:
\[
\begin{array}{rcl}
\varepsilon \circ (\hat g \times id_{D_x}) & = & g, \\
\varepsilon \circ (k \times id_{D_x})      & = & k,
\end{array}
\]
where $(\cdot \times \cdot)$ means a {\em functor product},
or, in the neutral to domains form,
\[
\begin{array}{rcl}
\varepsilon \circ <\hat g \circ Fst, Snd> & = & g, \\
\varepsilon \circ <k \circ Fst, Snd>      & = & k,
\end{array}
\]
where $<\cdot \circ Fst, Snd>$ is the same as
$(\cdot \times id_{D_x})$.

The notation is now wholly categorical and not so suitable.
The more sense is added by the language of {\em functors}.

\subsubsection{A system of types within cartesian closed category}

Now we give a brief sketch of viewing the cartesian closed
category (c.c.c.) as a system of types.

\begin{namelist}{{\tt -h} {\it {   }}}
\item {\hspace{-1em}\sf Theory of functions.} Each c.c.c represents
a {\em theory of functions}.
\item {\hspace{-1em}\sf Maps.} The {\em maps} in the category are
certain special functions that are used to express
the relations between the types (the domains of the category).
\item {\hspace{-1em}\sf Products.} In order to be able to deal with
multi-ary functions, we assume we can form and analyze
{\em products}.
\item {\hspace{-1em}\sf Function spaces.} In order to be able to work
with transformations of arbitrary functions
(arbitrary within the theory) we assume we can form
{\em function spaces}. Note, that the {\em higher} types
enter the theory, e.g., as the sequence of domains:
$$
D,\ (D \to D),\ ((D \to D) \to D), \dots\ .
$$
\item {\hspace{-1em}\sf Operations $\varepsilon$ and $\hat \cdot$.}
To be able
really to view these domains as function spaces, certain
operations, $\varepsilon$ and $\hat \cdot$, with characteristic
equations have to be laid down.
\item {\hspace{-1em}\sf Cartesian closed category.} C.c.c is a theory of
functions, and
the higher type functions are included. Hence, the theory
of c.c.c's is the {\em theory of types}. It is only one
such theory.
\item {\hspace{-1em}\sf `Bigger' theories.} `Bigger' theories could be
obtained by demanding more types, e.g., by axiomatizing
{\em coproducts} (disjoint sums) and $D_x + D_y$
\item {\hspace{-1em}\sf Type $[\ ]$.} We could throw in type $[\ ]$
of propositions so that higher types like $(D^n \to [\ ])$
correspond to $n$-ary predicates.
\end{namelist}

%%%%%%%%%%%%%%%%%%%%% Top of inclusion %%%%%%%%%%%%%%%%%%%%%%%%%%%
%\section{General requirements}\label{Section:2}

\section{Environment and an access to values}\label{Section:3}

To build the typed language we need to think of the values of the
variables. The values of the variables are available via {\em
access} functions from an environment $Env$. The representation of
an environment is given by the domains $D_y,\ D_x,\ \dots$ which
are ranges of possible values of $y,\ x,\ \dots$. The domains
$D_y,\ D_x$ give the {\em explicit} part of an environment $Env$,
and its {\em implicit} rest $E$, not be detailed for current
consideration, is separated from $D_y,\ D_x$:
$$
Env = (E \times D_y) \times D_x
$$

\subsection{Updating an environment}

Whenever we want to {\em update} $Env$ the restriction
is imposed to its counterparts:
$$
Env = Implicite\_part \times Explicite\_part,
$$
e.g.,
$$
Env = E \times D_x
$$
with `$E$' for implicit part and `$D_x$' for explicit part.

In fact, within $Env$ we have an `old' value of $x$ which ranges
$D_{x,Old}$, and $E$ which does not depend on $x$. The description
of updating an environment $Env$ has to include both its explicit
part $D_x$ and implicit rest $E$. An outline of {\em updating
process} is given below.
\begin{namelist}{{\tt -h} {\it {   }}}
\item {\hspace{-1em}\sf Step 1: Building the old environment.} To combine
$Env_{Old}$ we construe the product
$$
Env_{Old} = E \times D_{x,Old}.
$$
\item {\hspace{-1em}\sf Step 2: Bringing in a range of values.} The
product of $Env_{Old}$  and $D_x$ is generated as
$$
Env_{Old} \times D_x = (E \times D_{x,Old}) \times D_x.
$$
\item {\hspace{-1em}\sf Step 3: Establishing an $Update$-function.} The
An $Update_x$ function is established to enable the transformation
from $Env_{Old}$ to $Env_{New}$:
$$
Update_x : (E \times D_{x,Old})\times D_x \to E\times D{x,New},
$$
or,
$$
Update_x : Env_{Old} \times D_x \to Env_{New}.
$$
\end{namelist}
At this stage we are to compare the properties of the
domains $D_x$, $D_{x,Old}$, $D_{x,New}$:
\begin{itemize}
\item[(1)] $D_x$ is an unrestricted range for free variable $x$;
\item[(2)] $D_{x,Old}$ is some {\em existing}
(i.e., {\em before} evaluation) restriction of $D_x$;
\item[(3)] $D_{x,New}$ differs from $D_{x,Old}$ exclusively
in a point $x$.
\end{itemize}
All of this could be implemented in a particular kind of $Update_x$,
which would be referred as {\em substitution}, or $Subst_x$:
$$
Subst_x : Env_{Old} \times D_x \to Env_{New}.
$$
The description of its behavior by the elements gives the
following:
$$
Subst_x : [i,d] \longmapsto i_{(d/x)},
$$
where $d \in D_x$, $i \in Env_{Old}$, and $i_{(d/x)} \in Env_{New}$.
[Here: the new instance $i_{(d/x)}$ of environment is the same as
its old instance $i$ {\em excepting} the point $x$, which
is replaced by $d$.]

One could imagine that there is a pointer from `$x$' to its
possible values `$d$'.

\subsection{Viewing $Subst_x$ as a pointer}

Now we discuss the possibility to construe a pointer
to the partitions of an environment. At first, we would try
the equation
\[
\begin{array}{lcl}
Subst_x & = & Fst \times id_{D_x} \\
        & = & <Fst \circ Fst, id_{D_x} \circ Snd> \\
        & = & <Fst \circ Fst, Snd>,
\end{array}
\]
where $Fst \times id_{D_x}$ is a {\em functor product},
$<Fst \circ Fst, Snd>$ its linear notation, and
$$
Subst_x : Env_{Old} \times D_x \to Env_{New}
$$
for $Env = (E \times D_y) \times D_x$.

The functor product when being applied to {\em ordered pair}
generates an access separately to the {\em first}
and to the second its members. This feature makes it possible
to bring in the following maps as the pointers to the
partitions of the environment.
\begin{namelist}{{\tt -h} {\it {   }}}
\item {\hspace{-1em}\sf Pointer to the part independent on `$x$'.}
This is a composition of $Fst$'s which ranges the product $E
\times D_y$, i.e. and implicit -- and independent, -- part of the
environment:
$$
Fst \circ Fst : Env_{Old} \times D_x \to E \times D_y
$$
\item {\hspace{-1em}\sf Pointer to the part of new values for `$x$'.}
This is a second projection $Snd$ which ranges over the desired
domain $D_x$:
$$
Snd : Env_{Old} \times D_x \to D_{x,New}
$$
\item {\hspace{-1em}\sf Coupling the new environment.}
Now we generate an access to the new environment.
Taking into account the pointers for both the partitions,
we need to construe their {\em couple} to obtain the pointer
to the new environment:
$$
<Fst \circ Fst, Snd> : Env_{Old} \times D_x \to Env_{New}
$$
\end{namelist}

Getting started with a new environment
$$
Env_{New} = (E \times D_y) \times D_{x,New},
$$
we can evaluate the {\em arbitrary functions}. The process of
extracting the pointers to $D_y$, $D_{x,New}$ and generating the
values from $D'_y$ whenever $D_y = (D'_y)^{D_x} (= D'_y \to D_x)$,
i.e. for the {\em function space} $D_y$, comes down to the
following steps.
\begin{namelist}{{\tt -h} {\it {   }}}
\item {\hspace{-1em}\sf Step 1: Access to $D_y$.} We take
the first partition ($Fst$) of $Env_{New}$ and after that construe
the pointer to its second ($Snd$) partition:
$$
Snd \circ Fst : Env_{New} \to D_y
$$
\item {\hspace{-1em}\sf Step 2: Access to $D_{x.New}$.}
An effect of applying $Snd$ to $Env_{New}$ gives the pointer
$$
Snd : Env_{New} \to D_{x,New}
$$
\item {\hspace{-1em}\sf Step 3: Coupling an access to $D'_y$
for $D_y = (D'_y)^{D_x}$ by $\varepsilon$.} We take the
subpartitions of $Env_{New}$ as above and restore the pointer:
$$
<Snd \circ Fst, Snd> : Env_{New} \to ((D'_y)^{D_x} \times D_{x,New})
$$
taking in mind that
$\varepsilon : (D'_y)^{D_x} \times D_{x,New} \to D'_y$.
\end{namelist}
Now we are able to take a function $f$ from $(D'_y)^{D_x}$ and
the argument $d$ from $D_{x,New}$ and apply $f$ to $d$
using $\varepsilon$. Thus, the equation
$\varepsilon [f,d] = f(d)$ is valid giving rise to the
values $f(d)$ from $D'_y$

\subsection{Encapsulation of an object}

In particular, an evaluation process may result in
capturing the object being evaluated by an environment.

\begin{lem}[Citation]\label{Lemma:1}
For any given environment $Env = (E \times D_y) \times D_x$
and the domain $D_y = (D'_y)^{D_x}$
the {\em constant} $c \in D_x$ and
the {\em function constant} $f \in D_y$
are described by the maps $\hat {}(1_{\{c\}} \circ Snd)$
and $\hat {}(f \circ Snd)$ respectively.
\end{lem}
{\em Proof}. For any given instance $i \in Env$, e.g.,
$i = [[e,y],x]$ whenever $x, d \in D_x$, $y, f \in D_y$ then:

(1) $\hat {}(1_{\{c\}} \circ Snd)\ i\ c$ = $(1_{\{c\}} \circ Snd)[i, c]$ =
    $1_{\{c\}}c$ = $c$.

(2) $\hat {}(f \circ Snd)\ i\ d$ = $(f \circ Snd)[i,d]$ = $f(d)$.

Thus, this proof is straightforward and elementary.

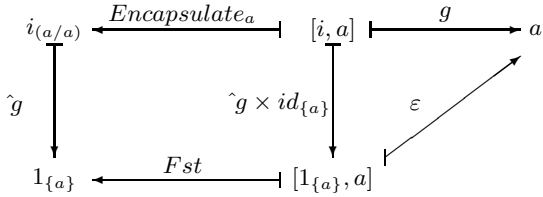
\begin{figure}
\unitlength 1.00 mm \special{em:linewidth 0.4pt}
\linethickness{0.4pt}
\begin{picture}(71.00,23.00)
\put(7.00,1.00){\makebox(0,0)[cc]{$1_{\{a\}}$}}
\put(37.00,1.00){\vector(-1,0){25.00}}
\put(44.00,1.00){\makebox(0,0)[cc]{$[1_{\{a\}},a]$}}
\put(37.00,21.00){\vector(-1,0){25.00}}
\put(24.00,23.00){\makebox(0,0)[cc]{$Encapsulate_a$}}
\put(24.00,3.00){\makebox(0,0)[cc]{$Fst$}}
\put(37.00,0.00){\line(0,1){2.00}}
\put(37.00,20.00){\line(0,1){2.00}}
\put(7.00,21.00){\makebox(0,0)[cc]{$i_{(a/a)}$}}
\put(7.00,19.00){\vector(0,-1){15.00}}
\put(2.00,11.00){\makebox(0,0)[cc]{$\hat {}g$}}
\put(44.00,21.00){\makebox(0,0)[cc]{$[i,a]$}}
\put(49.00,21.00){\vector(1,0){20.00}}
\put(51.00,4.00){\vector(4,3){18.00}}
\put(71.00,21.00){\makebox(0,0)[cc]{$a$}}
\put(55.00,11.00){\makebox(0,0)[cc]{$\varepsilon$}}
\put(44.00,19.00){\vector(0,-1){15.00}}
\put(37.00,11.00){\makebox(0,0)[cc]{$\hat {\ }g \times
id_{\{a\}}$}} \put(6.00,19.00){\line(1,0){2.00}}
\put(43.00,19.00){\line(1,0){2.00}}
\put(49.00,22.00){\line(0,-1){2.00}}
\put(57.33,-1.00){\line(0,0){0.00}}
\put(59.00,23.00){\makebox(0,0)[cc]{$g$}}
\put(51.00,5.00){\line(0,-1){2.00}}
\end{picture}
\caption{\sf Encapsulation of a constant $a$ (\small Notations and
explanation: $i$ is an instance of environment $Env$, thus
 $i \in Env$; $i_{(a/a)}$ means the instance of environment
 which captured the constant $a$, also means the
 substitution of domain for $a$ by $a$; a {\em closure}
 $1_{\{a\}}$ (identity map as a canonical evaluation)
 for the evaluated constant is generated; whenever a closure
 is not the identity map then the constant is not
 canonically evaluated. As may be shown, the map $g$
 is equal to $1_{\{a\}}\circ Snd$.)}\label{Diag:1}
\end{figure}

Canonical evaluation of a {\em constant} is according the
commutative diagram in Figure~\ref{Diag:1}. The reasons are as
follows. Let $i$ be an instance of environment $Env$, thus,  $i
\in Env$. Each occurrence of `$a$' canonically is replaced by the
same `$a$', i.e. $i_{(a/a)}$ means the instance of environment
 which captured the constant $a$, also means the
 substitution of domain for $a$ by $a$.
We need  a {\em closure} to trigger the evaluation
process, and
 $1_{\{a\}}$ (identity map as a canonical evaluation)
 for the evaluated constant is generated.
Roughly speaking, this identity map evaluates a constant and
 whenever a closure
 is not the identity map then the constant is not
 canonically evaluated.

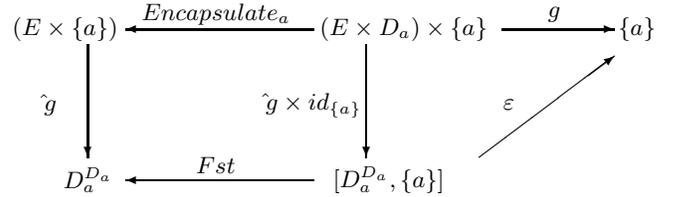
\begin{figure}
\unitlength 1.00 mm \special{em:linewidth 0.4pt}
\linethickness{0.4pt}
\begin{picture}(83.00,23.00)
\put(10.00,1.00){\makebox(0,0)[cc]{$D_a^{D_a}$}}
\put(40.00,1.00){\vector(-1,0){25.00}}
\put(50.00,1.00){\makebox(0,0)[cc]{$[D_a^{D_a},\{a\}]$}}
\put(40.00,21.00){\vector(-1,0){25.00}}
\put(27.00,23.00){\makebox(0,0)[cc]{$Encapsulate_a$}}
\put(27.00,3.00){\makebox(0,0)[cc]{$Fst$}}
\put(7.00,21.00){\makebox(0,0)[cc]{$(E\times\{a\})$}}
\put(10.00,19.00){\vector(0,-1){15.00}}
\put(5.00,11.00){\makebox(0,0)[cc]{$\hat {}g$}}
\put(52.00,21.00){\makebox(0,0)[cc]{$(E\times D_a)\times\{a\}$}}
\put(62.00,4.00){\vector(4,3){18.00}}
\put(83.00,21.00){\makebox(0,0)[cc]{$\{a\}$}}
\put(66.00,11.00){\makebox(0,0)[cc]{$\varepsilon$}}
\put(47.00,19.00){\vector(0,-1){15.00}}
\put(40.00,11.00){\makebox(0,0)[cc]{$\hat {}g \times id_{\{a\}}$}}
\put(54.33,-1.00){\line(0,0){0.00}}
\put(65.00,21.00){\vector(1,0){15.00}}
\put(72.00,23.00){\makebox(0,0)[cc]{$g$}}
\end{picture}
\caption{\sf Environment of encapsulation (\small Notations and
explanation: $D_a$ is a range of $a$-compatible objects, i.e.
those with the same type; for simplicity assume $D_a=\{a\}$;
singleton $\{a\}$ is the encapsulated constant. The map
$Encapsulate_a$ builds a renewed environment by setting up the
product of implicit partition of the environment with the
singleton $\{a\}$.) }\label{Diag:2}
\end{figure}

Now we describe the evolution of an environment when encapsulation
of the constant occurs. The environment in Figure~\ref{Diag:2} is
treated as the cartesian product of the range domains. The
notations naturally reflects the ranges, and $D_a$ is a range of
$a$-compatible objects, i.e. those with the same type. For
simplicity we assume $D_a=\{a\}$, and this singleton $\{a\}$ is
the encapsulated constant. The map $Encapsulate_a$ builds a
renewed environment by setting up the product of implicit
partition of the environment with the singleton $\{a\}$.

\subsection{Building a pointer to values}

\subsubsection{Evaluation of a variable}

For single {\em free variable} the element-wise reasons for
the evaluation are described by the commutative diagram
in Figure~\ref{Diag:3}.
To read this diagram we use the additional notations:
$d \in D_x$ for an {\em element} being substituted;
$1_{D_x}: D_x \to D_x$ for an {\em identity map}.

We try to `solve' this diagram relatively $g$ and $Subst_x$.

\begin{namelist}{{\tt -h} {\it {   }}}
\item {\hspace{-1em}\sf Solution for $g$.}
For every $i \in Env$ the maps
\[
\begin{array}{rllrll}
\hat {} g(i) & : & D_x \to D_x;   & \hat {} g(i) & : & d \mapsto d, \\
\hat {} g    & : & Env \to (D_x \to D_x);
                                  & \hat {} g  & : & i \mapsto 1_{D_x}, \\
 g           & : & Env \times D_x \to D_x;
                                  &  g         & : & [i,d] \mapsto d \\
\end{array}
\]
are valid, hence the following is a `solution':
$$
\fbox{$g = Snd$}.
$$
The value of a free variable is represented by an identity map.
Note that this diagram corresponds to some idea of {\em closure}:
free variable is supposed to be closed under the environment of its
evaluation.
\item {\hspace{-1em}\sf Solution for $Subst_x$.}
For every $i = [e,x]$ from `old' environment
the map $Subst_x$ gives $[e,d]$ as an instance of
`new' environment:
$$
Subst_x : [[e,x],d] \mapsto [e,d],
$$
hence,
$$
\fbox{$Subst_x = <Fst \circ Fst, Snd>$}.
$$
Therefore, the `solution' of diagram in Figure~\ref{Diag:3} for
$g$ and $Subst_x$ in case we evaluate a single free variable is
given by diagrams in Figure~\ref{Diag:4} and in Figure~\ref{Diag:5}.
\end{namelist}

\begin{figure}
\unitlength 1.00 mm \special{em:linewidth 0.4pt}
\linethickness{0.4pt}
\begin{picture}(71.00,23.00)
\put(7.00,1.00){\makebox(0,0)[cc]{$1_{D_x}$}}
\put(37.00,1.00){\vector(-1,0){25.00}}
\put(44.00,1.00){\makebox(0,0)[cc]{$[1_{D_x},d]$}}
\put(37.00,21.00){\vector(-1,0){25.00}}
\put(24.00,23.00){\makebox(0,0)[cc]{$Subst_x$}}
\put(24.00,3.00){\makebox(0,0)[cc]{$Fst$}}
\put(37.00,0.00){\line(0,1){2.00}}
\put(37.00,20.00){\line(0,1){2.00}}
\put(7.00,21.00){\makebox(0,0)[cc]{$i_{(d/x)}$}}
\put(7.00,19.00){\vector(0,-1){15.00}}
\put(2.00,11.00){\makebox(0,0)[cc]{$\hat {}g$}}
\put(44.00,21.00){\makebox(0,0)[cc]{$[i,d]$}}
\put(49.00,21.00){\vector(1,0){20.00}}
\put(51.00,4.00){\vector(4,3){18.00}}
\put(71.00,21.00){\makebox(0,0)[cc]{$d$}}
\put(55.00,11.00){\makebox(0,0)[cc]{$\varepsilon$}}
\put(44.00,19.00){\vector(0,-1){15.00}}
\put(37.00,11.00){\makebox(0,0)[cc]{$\hat {\ }g \times id_{D_x}$}}
\put(6.00,19.00){\line(1,0){2.00}}
\put(43.00,19.00){\line(1,0){2.00}}
\put(49.00,22.00){\line(0,-1){2.00}}
\put(57.33,-1.00){\line(0,0){0.00}}
\put(59.00,23.00){\makebox(0,0)[cc]{$g$}}
\put(51.00,5.00){\line(0,-1){2.00}}
\end{picture}
\caption{\sf Substitution of a variable (\small Notations and
explanation: $d \in D_x$ an element being substituted; $1_{D_x}:
D_x \to D_x$ is an identity map. In evaluation of a variable the
most important is its substitutional property. The {\em closure}
for a variable is generated resulting in its image $1_{D_x}$.)
}\label{Diag:3}
\end{figure}
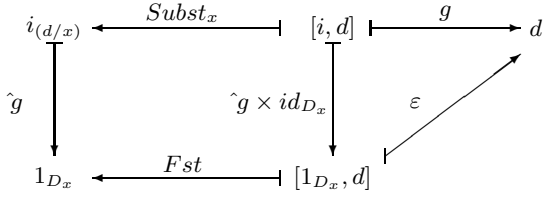

\begin{figure}
\unitlength 1.00 mm \special{em:linewidth 0.4pt}
\linethickness{0.4pt}
\begin{picture}(71.00,23.00)
\put(7.00,1.00){\makebox(0,0)[cc]{$1_{D_x}$}}
\put(37.00,1.00){\vector(-1,0){25.00}}
\put(44.00,1.00){\makebox(0,0)[cc]{$[1_{D_x},d]$}}
\put(37.00,21.00){\vector(-1,0){25.00}}
\put(24.00,23.00){\makebox(0,0)[cc]{$Fst \times id_{D_x}$}}
\put(24.00,3.00){\makebox(0,0)[cc]{$Fst$}}
\put(37.00,0.00){\line(0,1){2.00}}
\put(37.00,20.00){\line(0,1){2.00}}
\put(7.00,21.00){\makebox(0,0)[cc]{$i_{(d/x)}$}}
\put(7.00,19.00){\vector(0,-1){15.00}}
\put(2.00,11.00){\makebox(0,0)[cc]{$\hat {}Snd$}}
\put(44.00,21.00){\makebox(0,0)[cc]{$[i,d]$}}
\put(49.00,21.00){\vector(1,0){20.00}}
\put(51.00,4.00){\vector(4,3){18.00}}
\put(71.00,21.00){\makebox(0,0)[cc]{$d$}}
\put(55.00,11.00){\makebox(0,0)[cc]{$\varepsilon$}}
\put(44.00,19.00){\vector(0,-1){15.00}}
\put(35.00,11.00){\makebox(0,0)[cc]{$\hat { }Snd \times
id_{D_x}$}} \put(6.00,19.00){\line(1,0){2.00}}
\put(43.00,19.00){\line(1,0){2.00}}
\put(49.00,22.00){\line(0,-1){2.00}}
\put(57.33,-1.00){\line(0,0){0.00}}
\put(59.00,23.00){\makebox(0,0)[cc]{$Snd$}}
\put(51.00,5.00){\line(0,-1){2.00}}
\end{picture}
\caption{\sf Pointers for a variable (\small Notations and
explanation: an environment $i_{(d/x)}$ is the same as environment
$i$ excepted point $x$ which is replaced by $d$. In case of
evaluating a single free variable, map `$g$' from the diagram in
Figure~\ref{Diag:3} is to be replaced by the pointer
$Snd$.)}\label{Diag:4}
\end{figure}
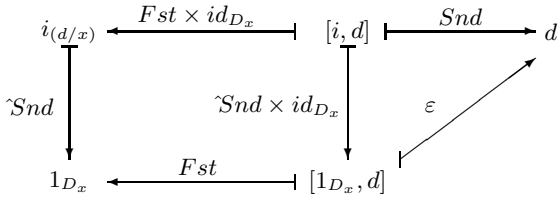

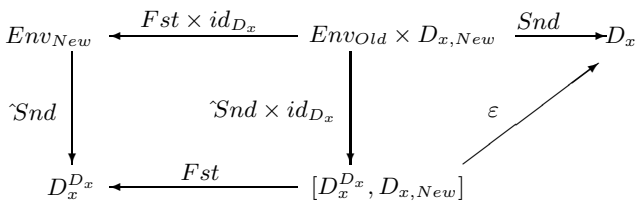
\begin{figure}
\unitlength 1.00 mm \special{em:linewidth 0.4pt}
\linethickness{0.4pt}
\begin{picture}(83.00,23.00)
\put(10.00,1.00){\makebox(0,0)[cc]{$D_x^{D_x}$}}
\put(40.00,1.00){\vector(-1,0){25.00}}
\put(52.00,1.00){\makebox(0,0)[cc]{$[D_x^{D_x},D_{x,New}]$}}
\put(40.00,21.00){\vector(-1,0){25.00}}
\put(27.00,23.00){\makebox(0,0)[cc]{$Fst \times id_{D_x}$}}
\put(27.00,3.00){\makebox(0,0)[cc]{$Fst$}}
\put(7.00,21.00){\makebox(0,0)[cc]{$Env_{New}$}}
\put(10.00,19.00){\vector(0,-1){15.00}}
\put(5.00,11.00){\makebox(0,0)[cc]{$\hat {}Snd$}}
\put(54.00,21.00){\makebox(0,0)[cc]{$Env_{Old}\times D_{x,New}$}}
\put(62.00,4.00){\vector(4,3){18.00}}
\put(83.00,21.00){\makebox(0,0)[cc]{$D_x$}}
\put(66.00,11.00){\makebox(0,0)[cc]{$\varepsilon$}}
\put(47.00,19.00){\vector(0,-1){15.00}}
\put(37.00,11.00){\makebox(0,0)[cc]{$\hat {}Snd \times id_{D_x}$}}
\put(54.33,-1.00){\line(0,0){0.00}}
\put(72.00,23.00){\makebox(0,0)[cc]{$Snd$}}
\put(69.00,21.00){\vector(1,0){12.00}}
\end{picture}
\caption{\sf Partitioning an environment $Env$ (\small
Explanation:  this diagram is expansion of the element-wise
commutative diagram in Figure~\ref{Diag:4} to the corresponding
domains.)}\label{Diag:5}
\end{figure}

\subsubsection{Evaluation of a constant function}

Evaluation of a {\em constant function} gives the most typical
sample to encapsulate the object of general nature. To observe the
effects we describe an applying of the constant function to the
argument. All the counterparts - both function and argument, -
from the category theory view are the {\em objects}.

The environment is changed whenever the application
of the function $f$ to the argument $x$ occurs,
i.e. the triggering event is $(fx)$, or similarly, $f(x)$.
In the environment an evaluation is triggered whenever the
value of argument `$d$' is passed to `$x$'.

The following Theorem~\ref{Theorem:1} reflects the
computational ideas in use.

\begin{th}[Citation of the function]\label{Theorem:1}

(1) The equation
\begin{center}
\begin{tabular}{p{3cm}cp{2.8cm}}
\(
f \circ
(\varepsilon \circ
           <\hat{} g_d \times id_{D_x}>
\)
           & = &
\(  \varepsilon \circ <\hat{} g_f \times id_{D_x}>
\)                                     \\
\end{tabular}
\end{center}
describes the object $f$ as a {\em functional constant}
parameterized by $g_d$ and $g_f$.

(2) The equation in (1) has the solution
\begin{center}
\begin{tabular}{lcl}
$g_d$ & = & $1_{D_x}\circ Snd$,  \\
$g_f$ & = & $f\circ Snd$,        \\
\end{tabular}
\end{center}
thus, the pointers to an environment
are generated.
\end{th}

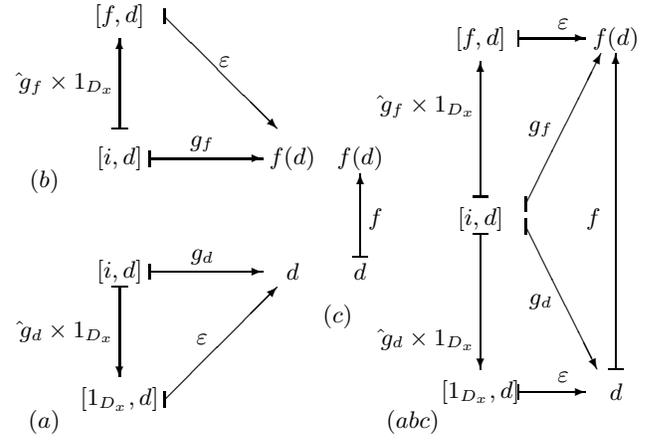
\begin{figure}
%\input p97_06.pic
%TexCad Options
%\grade{\off}
%\emlines{\off}
%\beziermacro{\on}
%\reduce{\on}
%\snapping{\on}
%\quality{2.00}
%\graddiff{0.01}
%\snapasp{1}
%\zoom{1.00}
\unitlength 1.00mm \linethickness{0.4pt}
\begin{picture}(82.00,57.00)
\put(15.00,22.00){\makebox(0,0)[cc]{$[i,d]$}}
\put(38.00,22.00){\makebox(0,0)[cc]{$d$}}
\put(26.00,24.00){\makebox(0,0)[cc]{$g_d$}}
\put(15.00,5.00){\makebox(0,0)[cc]{$[1_{D_x},d]$}}
\put(26.00,13.00){\makebox(0,0)[cc]{$\varepsilon$}}
\put(14.00,20.00){\line(1,0){2.00}}
\put(21.00,5.00){\vector(1,1){15.00}}
\put(21.00,6.00){\line(0,-1){2.00}}
\put(19.00,22.00){\vector(1,0){15.00}}
\put(19.00,23.00){\line(0,-1){2.00}}
\put(15.00,20.00){\vector(0,-1){12.00}}
\put(8.00,14.00){\makebox(0,0)[cc]{$\hat{}g_d \times 1_{D_x}$}}
\put(15.00,37.00){\makebox(0,0)[cc]{$[i,d]$}}
\put(38.00,37.00){\makebox(0,0)[cc]{$f(d)$}}
\put(26.00,39.00){\makebox(0,0)[cc]{$g_f$}}
\put(19.00,37.00){\vector(1,0){15.00}}
\put(19.00,38.00){\line(0,-1){2.00}}
\put(15.00,41.00){\vector(0,1){12.00}}
\put(21.00,56.00){\vector(1,-1){15.00}}
\put(14.00,41.00){\line(1,0){2.00}}
\put(21.00,55.00){\line(0,1){2.00}}
\put(8.00,47.00){\makebox(0,0)[cc]{$\hat{}g_f \times 1_{D_x}$}}
\put(15.00,56.00){\makebox(0,0)[cc]{$[f,d]$}}
\put(29.00,50.00){\makebox(0,0)[cc]{$\varepsilon$}}
\put(47.00,22.00){\makebox(0,0)[cc]{$d$}}
\put(47.00,37.00){\makebox(0,0)[cc]{$f(d)$}}
\put(46.00,24.00){\line(1,0){2.00}}
\put(49.00,29.00){\makebox(0,0)[cc]{$f$}}
\put(47.00,24.00){\vector(0,1){11.00}}
\put(63.00,29.00){\makebox(0,0)[cc]{$[i,d]$}}
\put(63.00,6.00){\makebox(0,0)[cc]{$[1_{D_x},d]$}}
\put(62.00,27.00){\line(1,0){2.00}}
\put(56.00,13.00){\makebox(0,0)[cc]{$\hat{}g_d \times 1_{D_x}$}}
\put(62.00,32.00){\line(1,0){2.00}}
\put(56.00,44.00){\makebox(0,0)[cc]{$\hat{}g_f \times 1_{D_x}$}}
\put(63.00,53.00){\makebox(0,0)[cc]{$[f,d]$}}
\put(68.00,5.00){\line(0,1){2.00}}
\put(68.00,52.00){\line(0,1){2.00}}
\put(69.00,30.00){\line(0,1){2.00}}
\put(69.00,27.00){\line(0,1){2.00}}
\put(74.00,55.00){\makebox(0,0)[cc]{$\varepsilon$}}
\put(74.00,8.00){\makebox(0,0)[cc]{$\varepsilon$}}
\put(63.00,32.00){\vector(0,1){18.00}}
\put(63.00,27.00){\vector(0,-1){18.00}}
\put(71.00,41.00){\makebox(0,0)[cc]{$g_f$}}
\put(71.00,18.00){\makebox(0,0)[cc]{$g_d$}}
\put(81.00,53.00){\makebox(0,0)[cc]{$f(d)$}}
\put(81.00,6.00){\makebox(0,0)[cc]{$d$}}
\put(81.00,9.00){\vector(0,1){42.00}}
\put(78.00,29.00){\makebox(0,0)[cc]{$f$}}
\put(68.00,53.00){\vector(1,0){9.00}}
\put(68.00,6.00){\vector(1,0){9.00}}
\put(69.00,28.00){\vector(1,-2){9.50}}
\put(69.00,31.00){\vector(1,2){10.00}}
\put(5.00,34.00){\makebox(0,0)[cc]{$(b)$}}
\put(5.00,2.00){\makebox(0,0)[cc]{$(a)$}}
\put(44.00,16.00){\makebox(0,0)[cc]{$(c)$}}
\put(54.00,2.00){\makebox(0,0)[cc]{$(abc)$}}
\put(80.00,9.00){\line(1,0){2.00}}
\end{picture}
\caption{\sf Evaluation of a constant function (\small
Explanation: this commutative diagram reflects a natural idea of
the {\em function constant}, i.e. the evaluated map does not
depend on an environment. A diagram to evaluate its {\em argument
variable} is similar to diagram in Figure~\ref{Diag:3}. The
diagram $(a)$ is a commutative-style description of evaluating the
{\em argument} of the function $f$. The diagram $(b)$ escribes the
intuitive reasons to observe $f$ as a {\em function constant}
which results in $f(d)$, the value of $f$ in a point $d$. The
diagram $(c)$ determines $f$ as the valid map. An assembling
diagrams $(a)$, $(b)$, and $(c)$ gives the commutative diagram
$(abc)$. All the diagrams contain the parameters $g_f$, $g_d$
which are to satisfy the {\em commutative} law. Thus, the
`solution' of diagram $(abc)$, if exists, relatively $g_f$, $g_d$
generates the {\em pointers} to access an
environment.)}\label{Diag:6}
\end{figure}

{\em Proof.} (1) The equation above is commented
as follows:
\begin{center}
\begin{tabular}{p{3cm}cp{2.8cm}}
Left part:  & & Right part: \\
\(
f \circ
(\underbrace{\varepsilon \circ
           <\hat{} g_d \times id_{D_x}>
            }
\)
           &  &
\(  \underbrace{\varepsilon \circ <\hat{} g_f \times id_{D_x}>
               }
\)                                     \\
eval of `$x$' within env `$i$'
when {\em actual parameter} `$d$'
is passed to argument `$x$'
           &  &
eval of `$fx$' within env `$i$'
when {\em actual parameter} `$d$'
is passed to argument `$x$'            \\
\end{tabular}
\end{center}
The premise of the sentence is described by the
commutative diagrams $(a)$, $(b)$, and $(c)$ in Figure~\ref{Diag:6}.
The equation is valid due to the existence of commutative diagram
$(abc)$, thus the conclusion is valid.

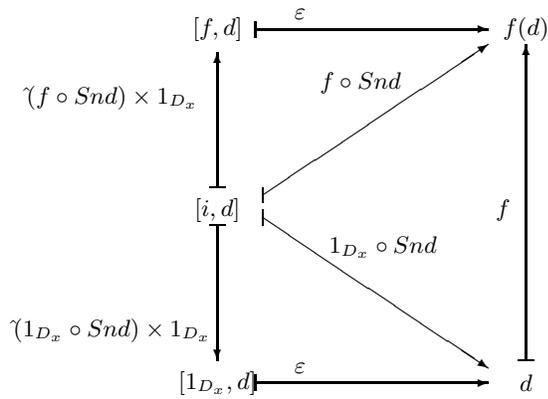
\begin{figure}
%\input p97_07.pic
%TexCad Options
%\grade{\off}
%\emlines{\off}
%\beziermacro{\on}
%\reduce{\on}
%\snapping{\on}
%\quality{2.00}
%\graddiff{0.01}
%\snapasp{1}
%\zoom{1.00}
\unitlength 1.00mm \linethickness{0.4pt}
\begin{picture}(76.00,55.00)
\put(34.00,29.00){\makebox(0,0)[cc]{$[i,d]$}}
\put(34.00,6.00){\makebox(0,0)[cc]{$[1_{D_x},d]$}}
\put(33.00,27.00){\line(1,0){2.00}}
\put(20.00,13.00){\makebox(0,0)[cc]{$\hat{}(1_{D_x}\circ Snd)
\times 1_{D_x}$}} \put(33.00,32.00){\line(1,0){2.00}}
\put(20.00,44.00){\makebox(0,0)[cc]{$\hat{}(f\circ Snd) \times
1_{D_x}$}} \put(34.00,53.00){\makebox(0,0)[cc]{$[f,d]$}}
\put(39.00,5.00){\line(0,1){2.00}}
\put(39.00,52.00){\line(0,1){2.00}}
\put(40.00,30.00){\line(0,1){2.00}}
\put(40.00,27.00){\line(0,1){2.00}}
\put(45.00,55.00){\makebox(0,0)[cc]{$\varepsilon$}}
\put(45.00,8.00){\makebox(0,0)[cc]{$\varepsilon$}}
\put(34.00,32.00){\vector(0,1){18.00}}
\put(34.00,27.00){\vector(0,-1){18.00}}
\put(53.00,46.00){\makebox(0,0)[cc]{$f\circ Snd$}}
\put(56.00,24.00){\makebox(0,0)[cc]{$1_{D_x}\circ Snd$}}
\put(75.00,53.00){\makebox(0,0)[cc]{$f(d)$}}
\put(75.00,6.00){\makebox(0,0)[cc]{$d$}}
\put(75.00,9.00){\vector(0,1){42.00}}
\put(72.00,29.00){\makebox(0,0)[cc]{$f$}}
\put(40.00,31.00){\vector(3,2){30.00}}
\put(40.00,28.00){\vector(3,-2){30.00}}
\put(39.00,53.00){\vector(1,0){31.00}}
\put(39.00,6.00){\vector(1,0){31.00}}
\put(74.00,9.00){\line(1,0){2.00}}
\end{picture}
\caption{\sf Pointers to access the environment with a constant
function (\small Explanation: this commutative diagram gives one
of the possible solutions of the diagrams in Figure~\ref{Diag:6}
relatively the parameters $g_f$ and $g_d$. Thus, the parameter
$g_d$ is replaced by $1_{D_x}\circ Snd$, $\hat{}g_d$ by
$\hat{}(1_{D_x}\circ Snd)$, $g_f$ by $f\circ Snd$, and $\hat{}g_f$
by $\hat{}(f\circ Snd)$.)}\label{Diag:7}
\end{figure}

(2) The existence of the pointers is due to Lemma~\ref{Lemma:1}.
Hence, the commutative diagram in Figure~\ref{Diag:7}
gives the needed pointers.

%%%%%%%%%%%%%%%%%%%%% Bottom of inclusion %%%%%%%%%%%%%%%%%%%%%%%%
\section{Conclusions}

A common object technique
equipped with the categorical and computational styles
is outlined.
As was shown, an object can be represented by
embedding in a host computational
environment. An embedded object is accessed by the laws
of the host system. A pre-embedded object is observed as
the decomposition into substitutional part and access function
part which are generated during the object evaluation.
They assist to easy extract of the result.

\section*{Acknowledgements}
The author is indebted to Institute for Contemporary Education
``JurInfoR-MSU'' for stimulating the research.

%%%%%%%%%%%%%%%%%%%%% Bibliography %%%%%%%%%%%%%%%%%%%%%%%%%%%%%%%
\addcontentsline{toc}{section}{References}

\newcommand{\noopsort}[1]{} \newcommand{\printfirst}[2]{#1}
  \newcommand{\singleletter}[1]{#1} \newcommand{\switchargs}[2]{#2#1}

%\bibliography{bbl_ad95,bibl_00,categ_97}
%%%%%%%%%%%%%%%%%%%%%%%%%%%%%%%%%%%%%%%%%%%%%
%
%\nocite{Cor:89} \nocite{Gil:90}
%
%\nocite{CaWe:85} \nocite{Ol:94} \nocite{AsGiNa:95} \nocite{KiCh:93}
%
%\nocite{Article:85:Elmasri:ERSetModel}
%\nocite{Article:89:Hoare:Summerschool}
%\nocite{Article:93:Johnson:AMAST93}
%\nocite{Article:94:Baclawski:CatTheoDatabase}
%\nocite{Article:94:Islam:CatTheoRM}
%\nocite{PhdThesis:94:Tuijn:Category}
%\nocite{Report:94:Lippe:CatTheory}
%\nocite{Report:94:Hofstede:CTPerspective}
%\nocite{Report:95:Hofstede:CT}

\end{document}